# Deep Generative Modeling for Mechanistic-based Learning and Design of Metamaterial Systems

Liwei Wang [a,b], Yu-Chin Chan[b], Faez Ahmed[b], Zhao Liu[c], Ping Zhu[a], Wei Chen[b]

a. The State Key Laboratory of Mechanical System and Vibration,
Shanghai Key Laboratory of Digital Manufacture for Thin-Walled Structures, School of Mechanical Engineering, Shanghai Jiao Tong University
Shanghai, P.R. China

b. Dept. of Mechanical Engineering, Northwestern University
Evanston, IL, USA

c. School of Design, Shanghai Jiao Tong University
Shanghai, P.R. China

## Abstract

Metamaterials are emerging as a new paradigmatic material system to render unprecedented and tailorable properties for a wide variety of engineering applications. However, the inverse design of metamaterial and its multiscale system is challenging due to high-dimensional topological design space, multiple local optima, and high computational cost. To address these hurdles, we propose a novel data-driven metamaterial design framework based on deep generative modeling. A variational autoencoder (VAE) and a regressor for property prediction are simultaneously trained on a large metamaterial database to map complex microstructures into a low-dimensional, continuous, and organized latent space. We show in this study that the latent space of VAE provides a distance metric to measure shape similarity, enable interpolation between microstructures and encode meaningful patterns of variation in geometries and properties. Based on these insights, systematic data-driven methods are proposed for the design of microstructure, graded family, and multiscale system. For microstructure design, the tuning of mechanical properties and complex manipulations of microstructures are easily achieved by simple vector operations in the latent space. The vector operation is further extended to generate metamaterial families with a controlled gradation of mechanical properties by searching on a constructed graph model. For multiscale metamaterial systems design, a diverse set of microstructures can be rapidly generated using VAE for target properties at different locations and then assembled by an efficient graph-based optimization method to ensure compatibility between adjacent microstructures. We demonstrate our framework by designing both functionally graded and heterogeneous metamaterial systems that achieve desired distortion behaviors.

# 1. Introduction

Metamaterials are artificial materials that derive their unusual properties from the geometry of microstructures rather than constituent materials [1]. A mechanical metamaterial is a subset of metamaterials designed to render a wide range of mechanical properties unreachable by its material composition [2]. As a result, there is a growing interest in designing multiscale mechanical metamaterial systems [3], assembled by numerous heterogeneous microstructures to achieve spatially varying macro-properties for intricate structural behaviors. However, the design of a multiscale metamaterial system is a challenging problem involving complex inverse design at the microscale, costly nested optimization at the macroscale, and boundary matching between neighboring microstructures.

Topology optimization (TO) methods consider the design of metamaterial microstructures as an optimization of the distribution of constituent materials within a periodically tiled unit cell. By using homogenization to compute the effective properties, the material distribution is updated iteratively to achieve target properties. However, current TO methods mainly focus on the design of extreme mechanical properties with volume fraction or mass constraints, such as negative Poisson's ratio design [4, 5] and the maximization of the bulk and shear modulus [6, 7], rather than the exploration of different microstructures to achieve wide-range material properties. This is because metamaterial design is an ill-defined inverse problem with an infinite-dimensional geometrical design space and a one-to-many mapping from properties to microstructures. These characteristics create an irregular landscape for the objective function with many local optima [8] and make microstructure design sensitive to the initial guess. When the full structure is assembled by heterogeneous rather than periodic microstructures, the design becomes even more challenging. State-of-the-art multiscale designs of systems with periodic microstructures [9-13] rely on nested optimization frameworks that are computationally demanding and do not scale to heterogeneous designs. Since the optimization of different microstructures is decoupled by the homogenization method, adjacent microstructures in a multiscale metamaterial system may not be well-connected. Even though some techniques have been proposed to address the connectivity issue [14-16], such as specifying fixed connectors and adding pair-wise constraints, they either sacrifice the generality of the design or do not scale well.

With the growth of data resources [17], a promising direction is to generate a database of the metamaterial microstructures and then utilize it to enable the efficient data-driven design of a heterogeneous system [18-27]. The spatial distribution of properties in the macrostructure can be first optimized using conventional topology optimization techniques and the corresponding microstructures can then be fetched from the database to fill each element in the full structure without the need to do nested optimization. To ensure connectivity, most data-driven methods only include a small number of microstructure designs with a narrow property space. For a wider range of properties, Schumacher et al. [28] proposed to construct a database with different metamaterial families, each containing similar microstructures and covering a respective region in the property space. However, this method relies on the elaborate construction of metamaterial families with limited variations, which can only provide suboptimal designs when one aims to explore the entire space of physical properties. To improve this method, Zhu et al. [8] established a larger and richer database by stochastically sampling the material distribution in the microstructure. However, due to the large amount and diverse shapes of microstructures, an immense combinatorial space needs to be explored to form compatible boundaries between neighboring unit cells. Overall, previous work has only focused on the construction of an elaborate database while few researchers have addressed how to incorporate a large database with the design of the multiscale system in a scalable way. The need for an effective representation and retrieval method for the metamaterials has been constantly overlooked.

The aim of this study is thus to provide an integrated framework for the representation, management, and utilization of a large microstructure database to facilitate metamaterial microstructures and multiscale systems design. Specifically, we will leverage the power of deep neural networks to discover the underlying data structure of a large database. Recent years have seen substantial advances in applying deep neural networks for structure and material designs, which can be largely divided into applications with predictive models and generative models. Predictive models aim to predict the response for a given design to lower the computational cost for the nested optimization, such as material properties prediction in multiscale structural design [29-32] and stress field prediction in topology optimization [33]. In contrast, deep generative models, such as generative adversarial networks (GAN) [34] and variational autoencoder (VAE) [35], aim to learn the underlying structure of a large dataset to enable the generation of new designs from a low-dimensional latent space. In the area of material design, deep generative models had been applied to the microstructure characterization and reconstruction of nanomaterials and alloys [36, 37], design of material microstructure morphologies [38], heat conduction materials [39], and design of photonic/phononic metamaterials [40-43]. Despite using different neural network architectures, these applications follow a similar design framework by using the latent vectors of the generative model as reduced-dimensional design variables for metamaterials. Combined with a trained predictive model, optimization on the latent space is performed to efficiently explore the high-dimensional or intractable geometric design spaces. However, the focus of these applications is on the microstructure design to achieve superior properties. How to incorporate the generative model into the data-driven design of multiscale metamaterial systems has not yet been explored.

In this study, we propose to train a deep generative model to organize complex microstructures into a continuous and highly structured latent space. Differentiating from existing deep-learning-based metamaterial design methods, our research highlights the neglected characteristics of the deep generative model: the regular mathematical structure and rich mechanical properties information encoded in the latent space. We show that complex manipulations on the geometry and tuning of the mechanical properties can be achieved by performing simple vector arithmetic in the latent space. By taking advantage of the discovered mathematical structure and mechanical properties information, the proposed generative model enables easy generation of metamaterial families with a controlled gradation of properties for multiscale functionally graded structure design and offers a scalable method to achieve heterogeneous multiscale system designs with optimal property distribution and compatible boundaries. We demonstrate our framework by designing both functionally graded and heterogeneous metamaterial systems that achieve desired distortion behaviors.

## 2. Overview of the proposed framework

We propose an integrated framework based on generative deep learning for the design of metamaterial microstructures, families, and multiscale metamaterial systems. As shown in Fig. 1, we first construct a generative neural network model and identify some important characteristics and mathematical structures. Based on the extracted characteristics, we propose to enable the tuning of properties, generation of diverse candidate microstructures for target properties, and metamaterial families with a prescribed gradation of properties by simple operations in the latent space. These design methods are integrated into the design of the aperiodic multiscale system as well as the functionally graded structure.

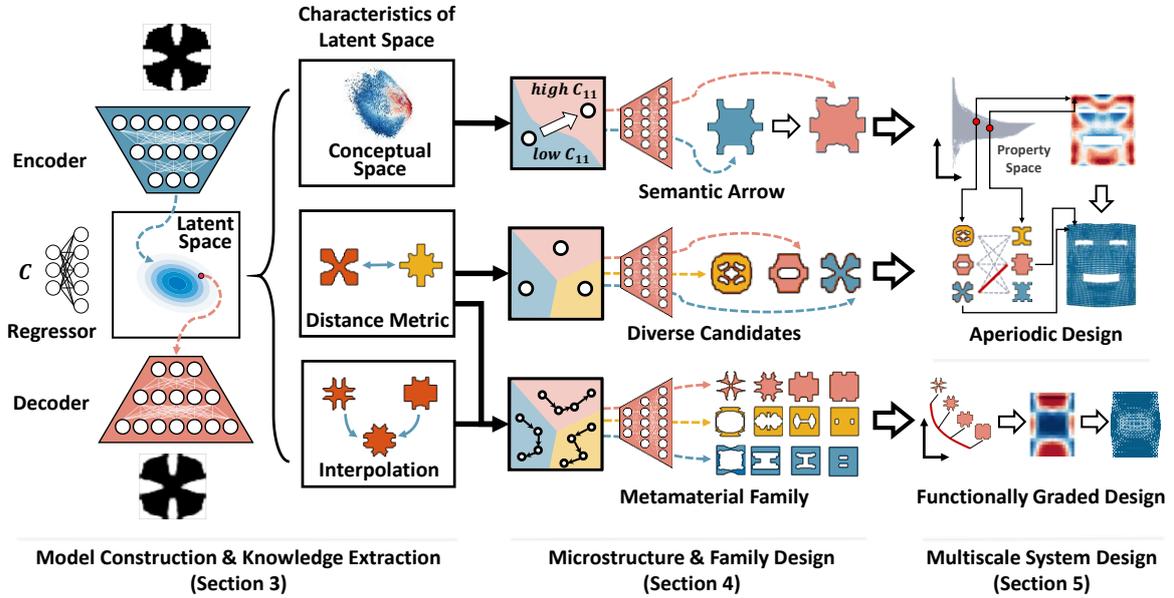

Fig. 1 Overview of the proposed framework

As shown in the first column of Fig.1, variational autoencoder (VAE) combined with a regressor of the mechanical properties of interest, i.e., the independent components of the stiffness matrix, are simultaneously trained on a large metamaterials database. This database contains various complex microstructures with wide-ranged properties precomputed by physics-based structure-property simulations. The encoder component of the VAE compresses complex microstructures into a low dimensional latent space by a series of convolutional operations while the decoder reconstructs the full microstructure from the compressed latent vectors through deconvolutions.

The low-dimensional latent space forms a bottleneck in the middle of the autoencoder, which will force the encoder to extract the most critical features of microstructures and organize complex geometries into a continuous and meaningful embedding. Since the compression and reconstruction processes only focus on the geometrical aspect, we incorporate a regressor into the architecture and train it simultaneously with the VAE model, organizing the distribution of microstructures in the latent space by their mechanical properties.

Due to the low dimensionality, the continuous embedding, and the relation with mechanical properties, the latent space is forced to encode meaningful variations in geometries or properties for certain directions in the latent space. The "meaningfulness" used in this study refers to the regular mathematical structure and rich mechanical properties information encoded in the latent space. Specifically, we observe that the latent space actually forms a *conceptual space* [44] where different abstract *concepts* of the microstructures, e.g., "*high stiffness*" and "*low Poisson's ratios*", occupy different regions in the space and can be represented by the latent vectors at the centers of those regions. The latent space also induces a distance metric to measure shape similarity and enable interpolation between a pair of microstructures by passing the weighted sum of their latent vectors to the decoder. Based on these characteristics, we can establish an integrated framework to facilitate the design of microstructures, metamaterial families, and multiscale structural systems.

For the unit-cell microstructure and microstructure family designs shown in the middle column of Fig. 1, we view the change from one concept to the other in the conceptual latent space to be like a *semantic operation*. For example, the process to make a microstructure change from "*low stiffness*" to "*high stiffness*" can be viewed as a semantic operation. The realization of a semantic operation generally involves complex shape transformations or iterative optimizations. However, we observe that complex semantic operations are achieved simply by moving in certain directions in the latent space, which are named as *semantic arrows* in this study [45]. We identify several such semantic arrows to enable the direct tuning of stiffness, anisotropic characteristics, and Poisson's ratio with simple vector arithmetic in the latent space. These semantic arrows provide high-level control of the microstructures, not only beneficial for unit cell design but also for generating a diverse candidate set to ensure connected neighboring microstructures and manufacturing feasibility in multiscale systems.

The advantageous characteristics of the latent space are further applied to the design of metamaterial families. Specifically, we define a metamaterial family to be a set of microstructures with similar configurations but a gradual change in geometry that also achieves a given continuous variation of material properties. In other words, it is a sequence of microstructures with morphing geometries sorted to render a certain gradation of material properties. Each metamaterial family can thus be represented as a bounded, continuous, and directed curve in the latent space, along which the corresponding microstructure will have a gradual change in geometry to achieve continuously graded properties. The metamaterial families in this study can be leveraged to design functionally graded materials (FGM) [46]. When we have prior knowledge on a specific design problem or would like to alleviate the stress concentrations, a metamaterial family instead of a large database is more desirable for the efficient design of the multiscale system. To create such families, we first view microstructures in the database as nodes on a graph with their mutual distance in the VAE latent space as edge weights. Different paths on this weighted graph connect different subsets of microstructures. The similarities of microstructures within each subset can be measured by the length of the corresponding path. Therefore, the shortest path on the graph connecting microstructures with the target gradation of properties can be considered as a piecewise linear approximation of the continuous curve representing a metamaterial family in the latent space. As a result, diverse sets of metamaterial families with the target property gradient can be efficiently obtained by searching for the shortest paths on this directed, weighted graph and then generating new microstructures through sampling on these paths.

Finally, as shown in the last column of Fig. 1, we integrate the microstructure and metamaterial family design methods into the design of aperiodic and functionally graded multiscale metamaterial systems, respectively. Both types of multiscale designs rely on a two-stage design process to achieve the prescribed system behavior. The first stage optimizes the properties distribution based on the property space provided by the large database or the given graded properties of a metamaterial family. The second stage selects corresponding microstructures for each designed property to assemble the full structure. In this stage for the aperiodic design, we propose to use a graph-based combinatorial optimization method to efficiently select the best microstructure from each candidate set to ensure good compatibility between adjacent unit cells. We provide detailed descriptions of the three main components of the proposed framework in Sections 3-5, respectively.

## 3. Knowledge extraction from variational autoencoder for metamaterials

### 3.1 Introduction of variational autoencoders

Current data-driven methods lack an efficient representation and data retrieval method that effectively enables the use of large metamaterial databases for multiscale system design. The recent development of deep generative models provides a new direction to address this issue. Deep generative models are neural networks that approximate high-dimensional observed data with simple sampling on the latent space [35]. The two popular deep generative models are the generative adversarial network (GAN) and the variational autoencoder (VAE). GAN is constructed based on a game theoretic scenario in which a generator network is competing against a discriminator network to generate new samples statistically indistinguishable from the observed data. In contrast, VAE simultaneously trains a parametric encoder in combination with a generator, providing an explicit likelihood for training. With the trained encoder, VAE allows an easy mapping from the observed dataset to the latent space with a continuous and meaningful coordinate system, which is generally intractable for GAN model. Although GAN is expected to have a greater generative ability and quality, it has a less explicit guarantee of a continuous latent space with a meaningful structure [47]. Since our aim is to provide a useful representation and management of microstructures for multiscale design, VAE is a more appropriate choice for this study.

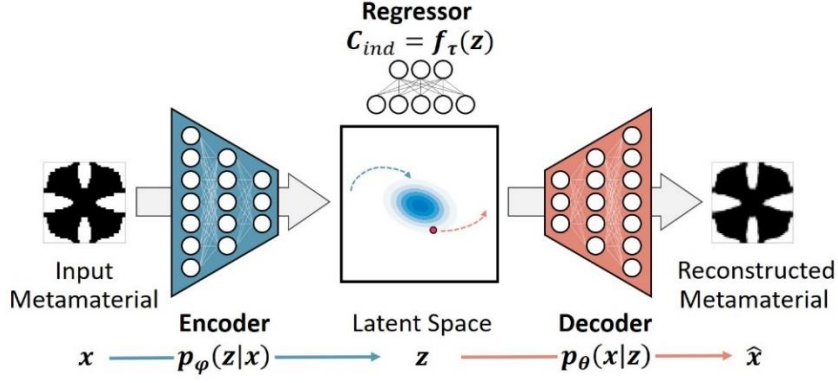

Fig. 2 Neural network architecture of the VAE model for metamaterials

For the metamaterial database, the pixelated matrix $x$ of microstructures can be viewed as a realization of an underlying random process with its true distribution $p^*(x)$ unknown. As shown in Fig. 2, VAE model aims to learn a stochastic mapping between the observed data space $x$ and a latent space $z$, which can be interpreted as a directed model with a joint distribution $p_\theta(x,z)$ over both the observed variables x and the latent variables z:

$$p_\theta(x,z) = p_\theta(x|z)p_\theta(z), \qquad (1)$$

where $\theta$ is the vector of model parameters, $p_\theta(z)$ is the prior distribution of latent variables and $p_\theta(x|z)$ is the approximated distribution of $x$ conditioned on $z$. The conditioned distribution $p_\theta(x|z)$ is parameterized by a deep neural network (*decoder*), which can provide almost arbitrary flexibility for the marginal distribution $p_\theta(x)$ with a relatively simple predefined prior distribution $p_\theta(z)$. However, the marginal distribution used to obtain the likelihood function for the training is generally intractable. To tackle this intractability issue, VAE introduces another deep neural network $q_\varphi(z|x)$ (*encoder* or *inference model*) to map $x$ back to the latent vector $z$ by approximating the posterior distribution $p_\theta(z|x)$. With the encoder and decoder networks, the likelihood function for the training has an explicit representation and is approximated by its *the evidence lower bound* (ELBO):

$$\mathcal{L}(\theta,\varphi;x) = E_{q_\varphi(z|x)}\left[\log\left(\frac{p_\theta(x,z)}{q_\varphi(z|x)}\right)\right]$$

$$= E_{q_\varphi(z|x)}[\log(p_\theta(x|z))] - E_{q_\varphi(z|x)}\left[\log\left(\frac{q_\varphi(z|x)}{p_\theta(z|x)}\right)\right]. \qquad (2)$$

Note that the second term is the Kullback-Leibler (KL) divergence $D_{KL}[q_\varphi(z|x) \| p_\theta(z|x)]$. This approximated lower bound enables the use of the efficient stochastic gradient descent method (SGD) for the simultaneous training of the encoder and decoder. Specifically, VAE assumes $p_\theta(z) \sim N(0, I)$ and adopts a Gaussian distribution for approximated posterior distribution:

$$q_\varphi(z|x) = N(\mu, \sigma), \tag{3}$$

where $\mu, \sigma$ is predicted by the decoder network. By assigning $z = \mu + \sigma \odot \varepsilon$ and $\varepsilon \sim N(0, I)$, the usual Monte Carlo estimator for ELBO used in SGD is reduced to

$$\mathcal{L}(\theta, \varphi; x) = E_{q_\varphi(z|x)}[log(p_\theta(x|z))] - D_{KL}[q_\varphi(z|x) \| p_\theta(z)]$$

$$= \frac{1}{L} \sum_l^L log(p_\theta(x|z)) - \frac{1}{2} \sum_j^J (1 + log(\sigma_j^2) - \sigma_j^2 - \mu_j^2) \tag{4}$$

where $J$ is the dimension of the latent space and $L$ is the number of samples. Therefore, we can optimize the parameters of the encoder and decoder by performing SGD to solve

$$\min_{\theta, \varphi} [-\mathcal{L}(\theta, \varphi; x)] \tag{5}$$

Besides the rigorous variational inference theory, we describe the VAE more intuitively as follows. Since the latent space generally has a much lower dimensionality compared to the original observed data, the VAE model forms an '*information bottleneck*' in the middle, greatly compressing the information of the original data into a low-dimensional latent space. The first term in Eq. (2) can be viewed as a reconstruction error that measures the loss of information due to the '*bottleneck*'. By minimizing this term, the encoder is forced to distill salient features of the observed data to enable the high reconstruction performance of the decoder. However, if trained only with the reconstruction loss, the model will tend to simply memorize all the training data and be reduced to a classical autoencoder, leaving many '*dead zones*' in the latent space that cannot be reconstructed into realistic samples by the decoder. Therefore, the sampling process during the training can be viewed as a way to improve the generalization ability by adding random noise to the latent variables, with the KL divergence added to the loss function for normalization. To ensure low reconstruction error under the random noise, the neighboring area in the latent space is mapped to similar microstructures by the decoder. Therefore, the encoder is forced to construct a continuous and meaningful latent space. In fact, a key insight of this paper is that the latent space of the VAE provides a natural distance metric and meaningful semantic structure, which enables easy control of complex geometries and an efficient data-driven design framework for metamaterial microstructures and multiscale system.

## 3.2 The proposed neural network architecture and model training

As illustrated in the last section, a stand-alone VAE learns a continuous latent space by only focusing on the geometries of microstructures. Since the interest of metamaterial design is to achieve prescribed mechanical properties, a latent space linked to the mechanical properties is more desirable. Therefore, as shown in Fig. 2 and 3, we augment the VAE architecture with an additional regressor, $f_\tau(z)$, whose inputs are the latent variables and outputs are the independent components $C_{ind}$ of the stiffness matrix $C$. The parameter optimization problem for the VAE is modified to include the regression error so that the model is trained simultaneously on geometries and properties:

$$\min_{\theta,\varphi,\tau}[-\mathcal{L}(\theta,\varphi;x) + \|f_\tau(z) - C_{ind}\|_2]. \tag{6}$$

The details of the neural network architecture are given in Fig.3. We compile convolutional layers to form the encoder and decoder. Note that the decoder is designed to have a shallower neural network structure than the encoder to reduce the possible overfitting. The dimension of the latent space is set to be 16 for this study, balancing between low-dimensionality and generation quality. Based on our empirical study, a higher dimensionality (>16) does not provide much improvement in reconstruction loss for our program. The regressor for the independent components of the stiffness matrix consists of layers of fully connected neural networks. During the training, the decoder will take the latent variables sampled from the approximated posterior distribution as input while the regressor will only take the mean value $\mu$ given by the encoder as input. In other words, the random noise is only added to the decoder but not the regressor.

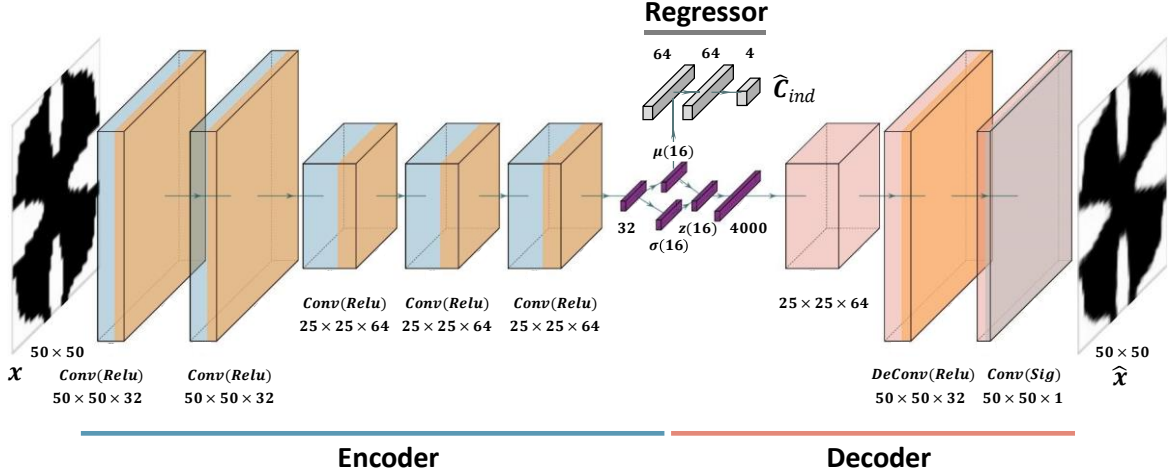

Fig. 3 The detailed architecture of the proposed neural network model

Before training, we construct a large database of microstructures as training data for the generative model, following our previously proposed method with a combination of topology optimization and an iterative stochastic shape perturbation [19]. Herein, we give a brief description of the database generation process. The Young's modulus and Poisson's ratio for the constituent material are set to be 1 and 0.49, respectively. In this study, to better demonstrate the benefit brought by the versatile properties of a large database, we aim to realize structural designs with spatially varying structural requirements, i.e. the target displacement profile design. Therefore, we focus on generating a large variation of geometries that cover a wide range of properties represented by the elements in the stiffness matrix. Since we only focus on the orthotropic microstructures in this study, four independent elements, i.e., $C_{11}$, $C_{12}$, $C_{22}$ and $C_{33}$ are used to describe the stiffness matrix. To cover a wide range of properties, we first performed SIMP-based TO [48] to find a corresponding pixelated microstructure design for each uniformly sampled target stiffness matrix. This TO design process is only performed once to form an initial database. After a threshold filtering, each microstructure design is represented by a 50×50 binary matrix. With 1400 microstructures generated by TO as initial seeds, an iterative stochastic shape perturbation algorithm is employed to perturb microstructure geometries that correspond to extreme and sparse properties. The extremeness and sparsity are measured by the distance to the boundary and the number of neighboring microstructures in the property space, respectively. Possible defects, such as isolated pixels and checkerboard patterns, are detected and fixed to ensure the feasibility of generated microstructures. By performing the selection and perturbation process for 200 iterations using parallel computing, we create a large database, close to 250,000 microstructures in our study, to systematically populate sparse areas and advance the boundary of the property space as shown in Fig. 4[1]. Based on our empirical study, this sample size not only provides an accurate deep learning model but also ensure compatibility between neighboring microstructures in the latter case studies of metamaterial systems design. Note that we only focus on the orthotropic microstructures in this study, which only requires four independent components, i.e., $C_{11}$, $C_{12}$, $C_{22}$ and $C_{33}$ to describe the stiffness matrix.

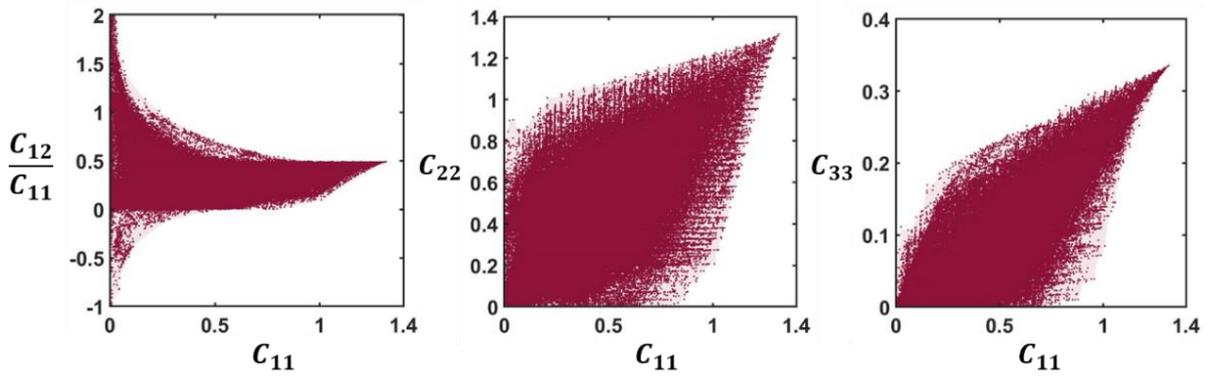

Fig. 4 Property space of the established metamaterial database, with shaded regions indicating the boundary of the property space.

For the training of the proposed neural network model, we randomly divide the database into a training set (228,396 microstructures) and a validation set (20,000 microstructures). The training is performed through mini-batch gradient descent for 500 epochs with the batch size set to be 32. We tried two commonly used optimizers, RMSprop and Adam, to train the model. The loss function converged to a lower value with RMSprop as the optimizer, taking 44,496 seconds on a workstation with a GeForce GTX Titan XP GPU and 12 GB memory. The training history of the loss function values is shown in Fig. 5.

---

[1] This database is available on our website. (https://ideal.mech.northwestern.edu/research/software/)

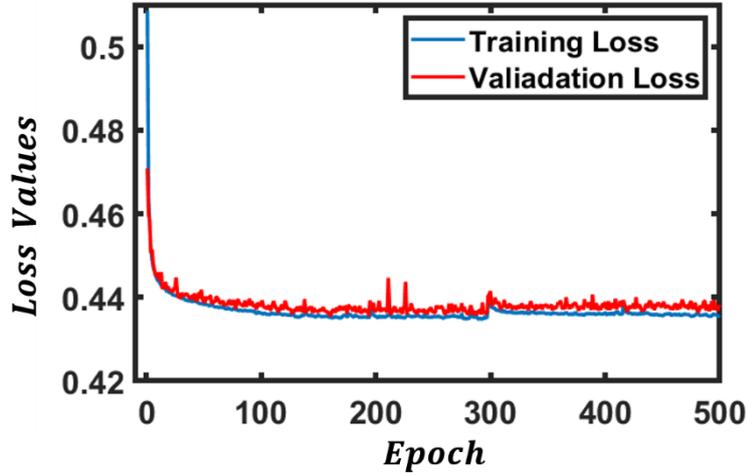

Fig. 5 The training history of the proposed neural network model

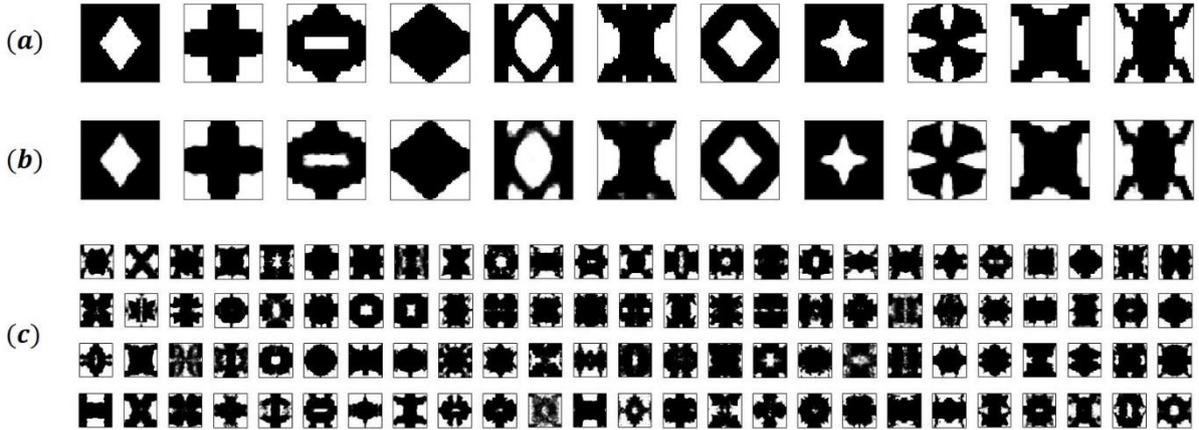

Fig. 6 Representative, reconstructed, and randomly generated microstructures, where black pixels are solid material and white pixels are void. (a) Representative ground truth microstructures selected from the database, (b) corresponding reconstructed microstructures generated by the proposed VAE model, (c) microstructures randomly generated with the proposed VAE model

Fig. 6(a) and (b) show some representative microstructure designs from the training data and their reconstructed configurations generated with the proposed model, respectively. Overall, the trained VAE model provides good reconstruction quality and can even preserve some fine features. However, it can also be noted that some reconstructed microstructures have blurred regions on the boundary, which is a common phenomenon for generative models with a log-likelihood loss function [47]. We randomly sample 100 points in the latent space and feed them into the decoder to obtain the generated designs shown in Fig.6(c). Despite a small portion of blurry designs, about one out of ten, most of the generated microstructures can have a clear and feasible configuration. Some newly generated microstructures do not exist within the training dataset and are observed to have diverse variations in geometries. This indicates that, as a generative model, VAE does not simply memorize the dataset but learns some underlying patterns that can be used to generate new and varied designs. To address the possible blurry designs, we interpret elements with the pixel values higher than 0.9 to be solid (mapped to 1) and others to be void (mapped to 0) in the remaining study. After filtering, morphological operations are used to detect and delete isolated pixels. Inevitably, this modification will cause discrepancies of latent vector and predicted properties between the original and modified microstructures. However, these discrepancies can be considered as insignificant in most cases if the samples are not in very sparse areas in the latent space, which is the case for the remaining parts in this paper. This is because blurred pixels generally occur on the boundary and the encoder is driven to map similar shapes into neighboring points in the latent space.

### 3.3 A deeper understanding of the latent space

In this section, we identify the important characteristics and underlying structures of the latent space. While these characteristics have been neglected in current metamaterial design methods, we leverage them for efficient representation and management of a large metamaterial database, which are the basic components of our proposed design framework.

When the continuity is combined with the low dimensionality of the latent space, different vectorized directions in the latent space encode physically meaningful patterns of microstructure shape morphing. To demonstrate this, we randomly choose a microstructure, map it to the latent space and then move each of the latent variables (16 in our study) independently along their negative and positive directions with a fixed step size. The corresponding microstructures along the path are then reconstructed with the decoder network. In Fig. 7(a), we show the results for traversals along the first three axes in the latent space ($z_1, z_2, z_3$ ).

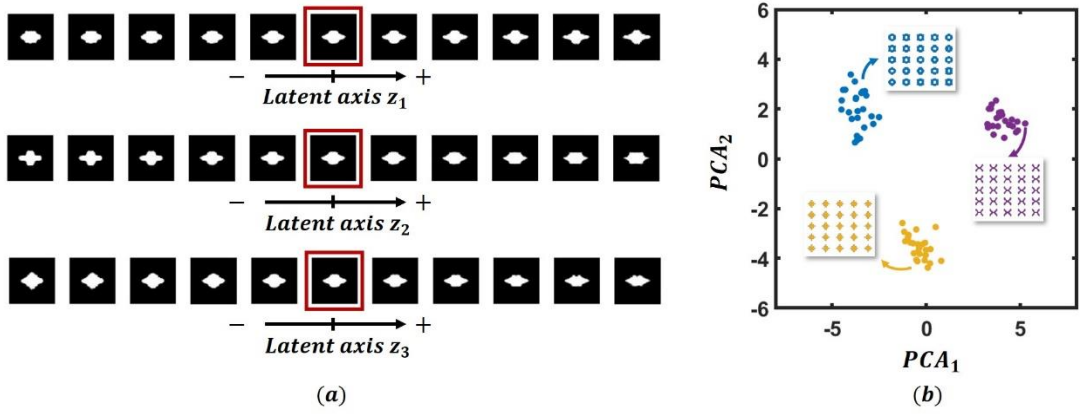

Fig. 7 (a) Generated microstructures by traversing along different axes in the latent space, and (b) shape clustering in the latent space using PCA.

It can be observed that each axis encodes a unique local feature transformation pattern for the selected microstructure. For example, by moving in the first axial direction, the central hole will be stretched in the horizontal direction. In contrast, the hole will be compressed in the vertical direction when moving along the second latent axis.

This meaningful structure of the latent space encompasses two characteristics that are critical to our later integrated design method. The first is that shape interpolation or morphing can be easily achieved by moving on a curve or line connecting the latent vectors of any two microstructures in the latent space and then mapping back to real microstructures using the decoder. This is illustrated in Fig. 7(a), where each traversal represents a natural morphing from the left-most microstructure to the rightmost one. The second characteristic rendered by this organized latent space is a natural distance metric between different shapes. A short transformation path in the latent space means that one microstructure can be changed into another shape with relatively simple transformation, providing a measure of the similarity between two shapes. To demonstrate this distance metric, we randomly select microstructures from three different locations in the latent space with a relatively large mutual distance. The two-dimensional principal component analysis (PCA) of these three clusters in the latent space and their corresponding microstructures are shown in Fig. 6(b). From the figure, each cluster contains microstructures with similar shapes while different clusters will have significantly different microstructure configurations, which is consistent with their mutual distance in the latent space.

The above characteristics of the latent space are all related to the geometrical aspect of microstructures. With the simultaneously trained regressor added to the VAE model, we further include mechanical property information into the latent space. For visualization, we reduce the 16-dimensional latent space of the whole database to a two-dimensional plane by PCA analysis and indicate the value of the mechanical property of each microstructure by color in Fig. 8.

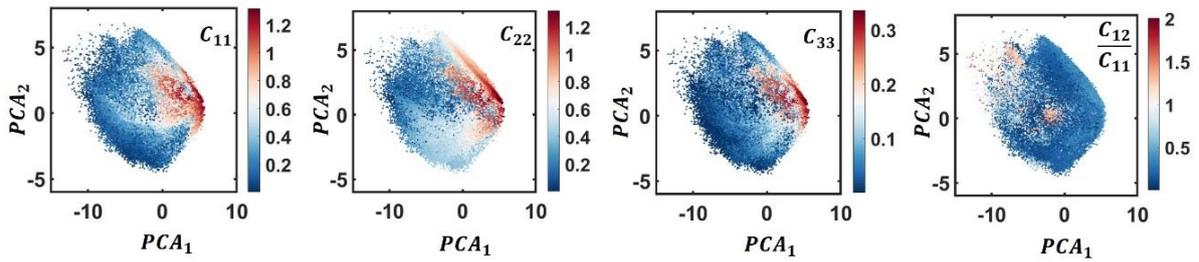

Fig. 8 Two-dimensional PCA analysis of the latent space colored by different properties. The color bar represents the value of a corresponding property.

From the figure, we note that microstructures with similar properties will cluster in the same areas in the latent space. This is because the regression loss will drive the model to adopt a latent embedding that can provide a relatively simple mapping between latent variables and the mechanical properties for a better regression performance. The latent space can be viewed as a geometrical representation of different design concepts for mechanical properties, e.g., "*high stiffness*" and "*low stiffness*", forming a conceptual space for metamaterials. This is the third characteristic that we found to be useful in our integrated design framework.

To summarize, the continuity and low dimensionality of the latent space leads to a meaningful and structured latent space. We identify three important characteristics of this latent space: a) interpolation can be achieved by moving in the latent space, 2) the distance in the latent space provides a measure for shape similarity, and 3) different property concepts cluster in the latent space to form a conceptual space. In the following sections, we will demonstrate how to take advantage of these characteristics to enable higher-level management of the microstructures and efficient design of its combinatorial multiscale systems.

## 4. VAE-assisted metamaterial microstructure and family design

### 4.1 Achieving mechanical properties using the concept of semantic arrows

We will demonstrate in this section how the deep generative models are used for inverse design of unit cell microstructures and a family of microstructures covering a range of properties, by performing simple vector arithmetic in the latent space.

As illustrated in Sec 3.3, the latent space of VAE forms a conceptual space for different design concepts related to mechanical properties. As a result, certain directions in this conceptual space represent a series of shape transformation to enable the change from one property to another, which are named semantic arrows in this study. Once we identify these semantic arrows, the tuning of the mechanical properties can be achieved by simple latent vector arithmetic without the need for direct and complex manipulations on high-dimensional shapes.

The first semantic arrow we obtain is related to the tuning of stiffness. It can be noted from Fig. 8 that microstructures with high and low $C_{11}$ values have different clusters in the latent space, indicating a semantic arrow for the transformation from low to high stiffness. To obtain this arrow, we first identify 30% of microstructures with the highest $C_{11}$ values and the other 30% with the lowest $C_{11}$ values to obtain their respective cluster center in the latent space. As shown in Fig. 9(a), the normalized vector pointing from the cluster center for low $C_{11}$ value to the one for high $C_{11}$ value can be obtained as a semantic arrow, which we name the $C_{11}$ *Arrow*.

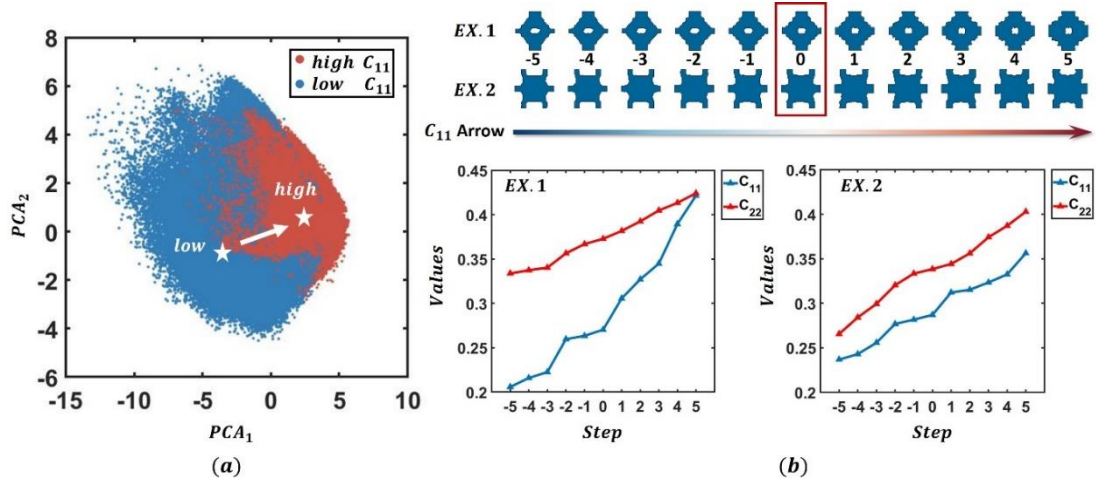

Fig. 9 The construction of the $C_{11}$ arrow and examples for the traversal along the $C_{11}$ arrow.

To illustrate the effect of traveling along this semantic arrow, we randomly select two microstructures and give them a set of negative and positive displacements along the $C_{11}$ arrow with a fixed step size. As shown in Fig. 9 (b), we use the decoder to reconstruct the corresponding microstructures after taking each step and calculate their effective $C_{11}$ values. By going in the positive direction of the modulus arrow, the microstructures will change in a way that increases the $C_{11}$ values and vice versa.

Note that the $C_{22}$ values will have the same increasing trend when moving along the $C_{11}$ arrow. However, we can also notice that there are some clusters with distinct $C_{11}$ and $C_{22}$ values from Fig. 8, indicating that a corresponding semantic arrow exists. Therefore, we first identify two sets of microstructures with $C_{11}/C_{22} > 2$ and $C_{22}/C_{11} > 2$, respectively, to represent two different anisotropic characteristics, i.e., stiffer in the first or the second axial directions. As shown in Fig. 10(a), the normalized vector between the centers of these two sets of microstructures in the latent space is obtained as an *Anisotropic Arrow* by simple vector arithmetic. Using the same representative microstructures, we reconstructed the corresponding microstructures by stepping along the negative and positive direction of the *Anisotropic Arrow*, which is shown in Fig. 10(b).

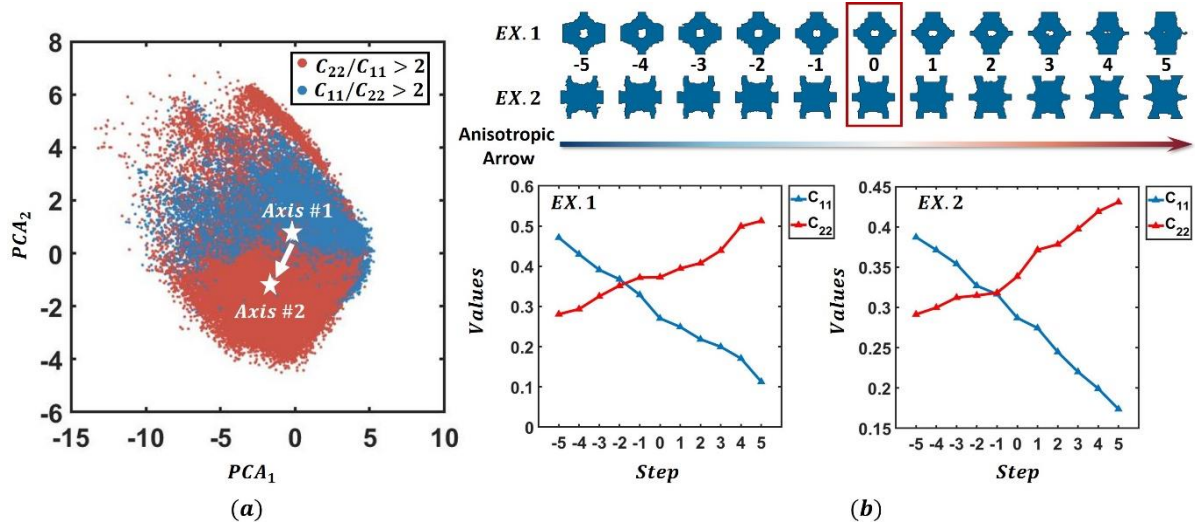

Fig. 10 The construction of the anisotropic arrow and examples for the traversal along the anisotropic arrow.

From the figure, it can be noted that the microstructures will expand in the vertical axis and shrink in the horizontal axis when moving along the *Anisotropic Arrow*. As a result, $C_{22}$ values gradually increase and surpass the $C_{11}$ values, changing the principal load-resisting direction from the first axis to the second one. Besides the examples above, there are other semantic arrows in the conceptual space, e.g. semantic arrows for Poisson's ratio and $C_{33}$, which can be indicated from different clusters in Fig. 8.

These semantic arrows demonstrate that the latent space of VAE forms a conceptual map with different directions representing different semantic operations. The complex operations on the configurations or the mechanical properties can be easily achieved by performing simple vector arithmetic in the latent space. This provides a high-level control and exploration method of abstract or complex attributes without the need to directly process high-dimensional shape representations.

### 4.2 Diverse microstructure selection and generation

One major advantage of data-driven design with a large microstructure database is that it can provide multiple candidates to achieve the same mechanical properties. From the candidate sets, the best-matched neighboring microstructures can be selected for the multiscale system design to achieve compatible connections. To avoid an immense combinatorial design space, the candidate set for a target property should be relatively small. However, a smaller candidate set will have a higher chance to be dominated by similar microstructures, which provide fewer connecting choices for the adjacent microstructures and thus decrease compatibility. Therefore, obtaining a small but diverse candidate set for a given property is critical for efficient multiscale system design.

Since the latent space of the VAE provides a natural distance metric between microstructures, we propose to first cluster microstructures with target properties in the latent space and then select one optimal candidate from each cluster to obtain a small but diverse candidate set. Specifically, k-means clustering is performed on the first two principal components of the latent variables to obtained different clusters. To demonstrate the effect of clustering, we set two target properties as examples and select a candidate set with 10 microstructures for each target by the proposed method, as shown in Fig. 11. As a comparison, we also include a candidate set constructed by greedily selecting ten most optimal microstructures in the database.

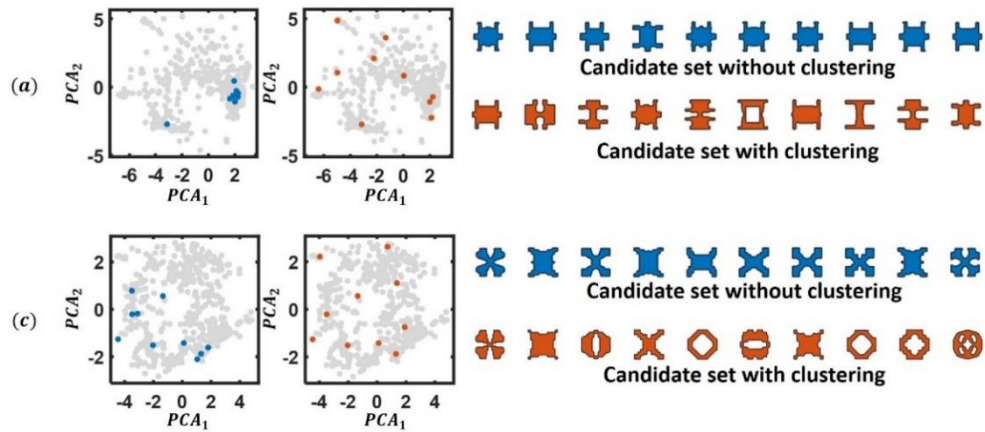

Fig. 11 Candidate sets obtained with/without clustering and PCA analysis of their latent spaces

Due to the dense database, the properties of these candidate sets are close to the objective with MSE values less than 0.01. However, candidate sets without clustering tend to include microstructures with similar configurations. In contrast, candidate sets selected from different clusters in the latent space will include relatively diversified shapes. This can also be indicated from their distributions in the latent space, where the diversified candidate sets cover a larger latent space and distribute more uniformly. Suppose two horizontally neighboring microstructures in the full structure are assigned with these two objective properties shown above. Without the clustering, only by selecting the fourth microstructure for the first objective property can we have the chance to obtain a connecting pair from the two candidate sets shown in Fig. 11. In contrast, by using clustering in the latent space, more choices are provided to achieve better compatibility. This further illustrates why a diverse candidate set is more desirable in the multiscale system design.

Besides passively selecting existing microstructures from the database, the proposed candidate set selection method can also provide a diversified initial guess for the TO method to achieve properties that are relatively extreme or uncommon in the database. The rationale behind this is that finding a corresponding microstructure for a given property is an inverse problem with infinitely many solutions, making gradient-based TO methods sensitive to the initial guess [8]. By selecting a diverse candidate dataset with properties close to the target as initial guesses, TO has a higher chance of converging to a local optimum with better performance and can efficiently explore different solutions. The proposed methods to obtain a diverse candidate set for a given target property will be integrated into the later multiscale systems design.

### 4.3 Metamaterial family design

The previous subsections focus on applying the generative model to assist the design at the unit-cell microstructure level. In this subsection, we turn to the design at the family level and propose an efficient method to generate metamaterial families by taking advantage of the large database and its highly structured latent space. On one hand, a metamaterial family can be used to construct a reduced database for a more efficient data-driven optimization when we have prior knowledge on a specific design problem. On the other hand, a controlled gradation of the mechanical properties can alleviate incompatibility between neighboring microstructures and thus decrease stress concentrations while achieving spatially varying properties for a specific function.

A controlled gradation of the mechanical properties can be considered as a hyper curve in the microstructures' property space. For a given gradation, we collect all microstructures in the database with their distance to the corresponding hyper curve smaller than a threshold $\delta$. These selected microstructures are sorted in ascending order by one of the mechanical properties to be controlled, forming a sorted sequence $\mathcal{H}$. To obtain metamaterial families with a gradual change of shapes, we first construct a directed weighted graph $G$ by taking each selected microstructure as a node, connecting to its $k$-nearest microstructures with higher ranks in the sorted sequence $\mathcal{H}$. Each edge points from low-rank to high-rank microstructures in the sorted sequence $\mathcal{H}$, taking the mutual distance in the latent space as the edge weight to measure the shape similarity between the two nodes. We further add two artificial nodes, i.e., the source node $s$ and the destination node $d$, into the weighted directed graph $G$, connecting the first and last $N$ nodes in the sorted sequence, respectively. With this, to find a metamaterial family with a controlled gradation in properties can be transformed into the searching for the shortest path connecting $s$ and $d$ on $G$. This can be achieved efficiently by mature graph search algorithms. Herein, we use one of the most used algorithms, the Dijkstra algorithm[49], to obtain the shortest path. Different metamaterial families with the same controlled gradation can be obtained by sequentially performing several runs of shortest path search. The nodes existing in all previously obtained paths will be deleted for each search.

To validate our method, we define the following property gradation as the design target:

$$C_{22} = C_{11},$$

$$C_{12} = \left[(1 - v^M) \cdot \left(1 - \frac{C_{11}}{C_{11}^M}\right)^4 + v^M\right] C_{11}, \qquad (7)$$

$$C_{33} = 0.25 \cdot (C_{11})^3 - 0.65 \cdot (C_{11})^2 + 0.6775 \cdot C_{11}$$

where $C_{11} \in [0, C_{11}^M]$, $C_{11}^M$ and $v^M$ are the $C_{11}$ and Poisson's ratio of the constituent material. In the constructed graph, $k$ for the $k$-nearest search is set to be five while the number of nodes $N$ connecting to the source $s$ and destination node $d$ is 50. The search process is performed five times to generate five different metamaterial families. The shortest paths obtained on the graph and corresponding microstructures are shown in Fig.12.

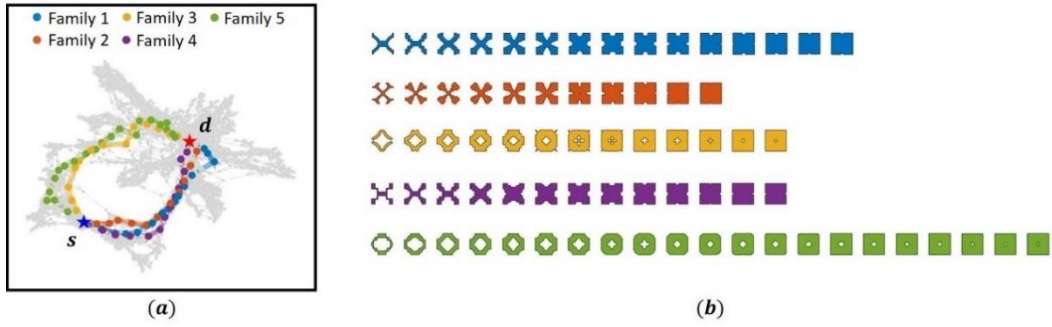

Fig. 12 Generated metamaterial families with the controlled gradation of properties. (a) The weighted directed graph constructed on the latent space, with different shortest paths marked with different colors. (b) Generated metamaterial families

From the graph shown in Fig. 12 (a), we can note that the shortest paths found contain different numbers of nodes, which correspond to metamaterial families with a different number of microstructures shown in Fig. 12(b). While different metamaterial families have relatively different patterns for microstructures, the transition of shapes within each family is smooth. The properties of these generated metamaterial families are shown in Fig. 13 (a).

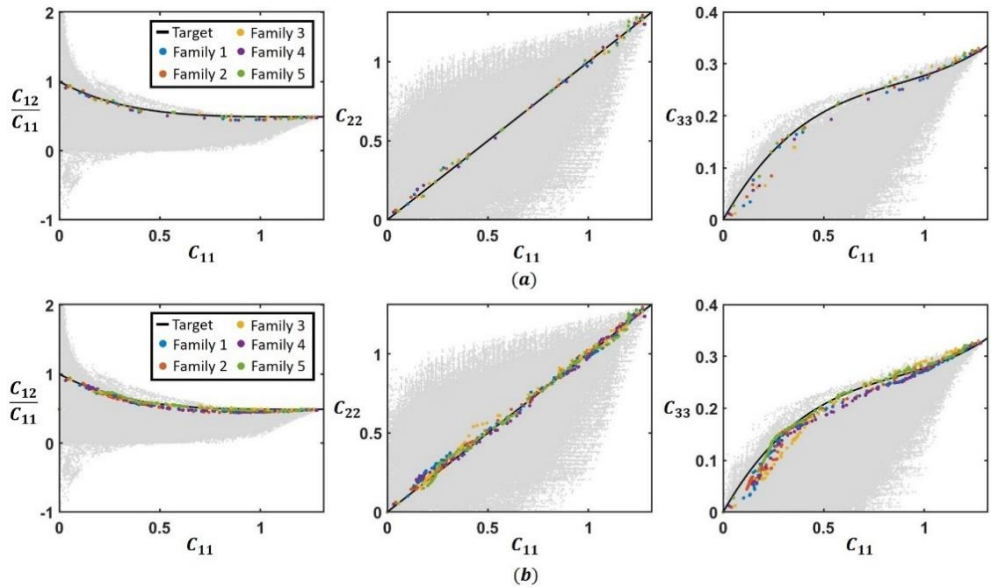

Fig. 13 Properties of the generated metamaterial families (a) Properties of the generated metamaterial families without interpolation (b) Properties of the generated metamaterial families with interpolation

As shown in the property space, all these metamaterial families follow the target gradation precisely. The generative model also enables the interpolation between microstructures within each family to generate a continuous metamaterial family. To demonstrate this, we perform uniform interpolation and extrapolation in the latent space for each metamaterial family and then pass these new latent vectors to the decoder for the generation of microstructures. The properties of these newly generated microstructures are evaluated and shown in Fig. 13(b). These new microstructures still accurately satisfy the target gradation of properties. Therefore, a set of continuously changing metamaterials can be generated by an efficient graph search on the latent space without the need to perform any direct manipulations or gradient-based optimization on high-dimensional representations of microstructures. This provides greater freedom and efficiency for the metamaterial family design.

## 5. VAE-assisted multiscale metamaterial system design

### 5.1 Proposed framework for multiscale system design

By integrating the design methods in the previous sections, we propose a two-stage framework for the data-driven design of a multiscale system with spatially varying properties to achieve a target structural behavior. Specifically, the first stage is to optimize the macro-property distribution while the second stage utilizes the optimized property distribution to select corresponding microstructure designs and assemble a full structure with compatible boundaries for neighboring microstructures. While this framework has a broader impact on multiscale metamaterial systems design, we demonstrate its benefit by designing structures with prescribed distortion behavior in this study.

In the macro-property optimization stage, the spatial properties distribution is optimized based on the properties space of the large metamaterial dataset or a given metamaterial family. We discretize the design area into four-node quadrilateral finite elements with the same size and define the optimization problem to be:

$$\min_{C_e} \|\gamma \odot u - u_t\|_2^2$$
$$s.t. K(C_e)u = f \tag{8}$$

$$\varphi_e(\boldsymbol{C}_e) \geq 0$$
$$\tau_e(\boldsymbol{C}_e) = 0,$$

where $\boldsymbol{C}_e$ is the stiffness matrix for the microstructure assembled in the element $e$, $\boldsymbol{\gamma}$ is a vector with the value 1 at the degrees of freedom corresponding to the displacements of interest and with zeros at all other entries, $\odot$ represent the element-wise product, $\boldsymbol{u}$ and $\boldsymbol{f}$ are the displacement and load vectors respectively, $\boldsymbol{u}_t$ is the target displacement vector with the specified values at the degree of freedom corresponding to the displacement of interest and with zeros at all other entries, $\varphi$ is the inequality constraint to force feasible, positive-valued properties, $\tau$ is the equality constraint on the properties and $\boldsymbol{K}$ is the global stiffness matrix depending on $\boldsymbol{C}_e$ of element $e$:

$$\boldsymbol{K} = \sum_{e=1}^{N_e} \widetilde{\boldsymbol{K}}_e(\boldsymbol{C}_e) \tag{9}$$

$$\boldsymbol{K}_e = \boldsymbol{B}^T \boldsymbol{C}_e \boldsymbol{B}, \tag{10}$$

where $N_e$ is the number of elements, $\widetilde{\boldsymbol{K}}_e$ and $\boldsymbol{K}_e$ are the element stiffness matrix in the global and elementary level respectively, and $\boldsymbol{B}$ is the constant gradient matrix for the four-node element. For the multiscale system design with a given metamaterial family, we only have the equality constraint $\tau$ on the properties to ensure the given property gradients. For the design with the whole database, the inequality constraint $\varphi$ will replace $\tau$ to ensure the feasibility of the optimized properties. Although there is no strict theoretical bound on the feasible properties, the rich and dense database we constructed provides a clear boundary on the achievable properties of the existing microstructures. Specifically, the value of signed L2 distance field to the boundary is calculated for each node on a Cartesian grid enclosing the current properties space composed by $C_{11}$, $C_{12}$, $C_{22}$ and $C_{33}$. The signed L2 distance field within the grid can then be estimated by interpolation. The feasibility of a given property can be indicated from this signed L2 distance field, with positive and negative values to indicate feasible and infeasible properties, respectively. To enable the use of a gradient-based optimization method, we calculate the partial derivatives of the signed L2 distance field with respect to property for each node by the finite difference method. This calculation is only performed once right after the construction of the database, which will not add extra computational cost to the optimization process. During the optimization, the partial derivatives of the property constraint can be directly obtained by simple interpolation without the need to do partial derivatives on-the-fly. It should be noted that an alternative to handle the property constraint is to use a surrogate model to explicitly approximate the signed L2 distance field and then obtain its gradient by direct differentiation. This method can be directly incorporated into the current framework without any modification. However, it is not trivial to obtain a smooth surrogate model that can precisely describe the highly irregular signed L2 distance field of the property space, which is beyond the scope of this study.

Since the constraint on properties is defined for each element, the overall number of constraints will be immense, resulting in an extremely time-consuming sensitivity analysis process. Therefore, we aggregate the elemental constraints into a single global constraint through Heaviside projection-based integral (HPI) [50, 51], transforming the original optimization problem into:

$$\min_{C_e} \mathcal{F} = \|\boldsymbol{\gamma} \odot \boldsymbol{u} - \boldsymbol{u}_t\|_2^2$$

$$s.t. \boldsymbol{K}(\boldsymbol{C}_e)\boldsymbol{u} = \boldsymbol{f}$$

$$C_{ij,min} \leq C_{e,ij} \leq C_{ij,max} \tag{11}$$

$$\frac{1}{N_e}\sum_{e=1}^{N_e} S\big(-\varphi_e(\boldsymbol{C}_e)\big) \leq \frac{1}{N_e}$$

where $C_{ij,max}$ and $C_{ij,min}$ are the maximal and minimal values for $C_{ij}$ in the database, respectively, $S$ is a function deified to be

$$S(x) = \tfrac{1}{2}(\tanh(\beta x) + 1), \tag{12}$$

which is an approximated but continuous version of the Heaviside step function with $\beta$ to be a given positive constant. With this problem definition, the sensitivity needed for optimization can be derived by the adjoint method:

$$\frac{\partial \mathcal{F}}{\partial C_{e,ij}} = -2[\boldsymbol{\gamma} \odot (\boldsymbol{u} - \boldsymbol{u}_t)]^T \boldsymbol{K}^{-1} \frac{\partial \widetilde{\boldsymbol{K}}_e}{\partial C_{e,ij}} \boldsymbol{u}, \tag{13}$$

where $\frac{\partial \widetilde{\boldsymbol{K}}_e}{\partial C_{e,ij}}$ can be transformed from the element level expressions:

$$\frac{\partial \boldsymbol{K}_e}{\partial C_{e,ij}} = \boldsymbol{B}^T \frac{\partial \boldsymbol{C}_e}{\partial C_{e,ij}} \boldsymbol{B}. \tag{14}$$

The detail derivation is given in Appendix A. In the real implementation, we only take the independent non-zero components of the stiffness matrix as design variables, i.e., $C_{11}$, $C_{12}$, $C_{22}$ and $C_{33}$. Therefore, the sensitivity value should be modified to be

$$\frac{\partial \mathcal{F}}{\partial C_{e,ij}} = \begin{cases} -2[\boldsymbol{\gamma} \odot (\boldsymbol{u} - \boldsymbol{u}_t)]^T \boldsymbol{K}^{-1} \frac{\partial \widetilde{\boldsymbol{K}}_e}{\partial C_{e,ij}} \boldsymbol{u}, & i = j = 1,2,3 \\ -4[\boldsymbol{\gamma} \odot (\boldsymbol{u} - \boldsymbol{u}_t)]^T \boldsymbol{K}^{-1} \frac{\partial \widetilde{\boldsymbol{K}}_e}{\partial C_{e,ij}} \boldsymbol{u}, & i = 1, j = 2 \end{cases} \quad (15)$$

These sensitivity values can then be passed to the MMA algorithm [52] to obtain the optimized properties distribution by solving (11).

In the microstructure assembly stage, we need to select or generate corresponding metamaterial microstructures to achieve the optimized properties assigned for each element in the macro-property optimization stage to assemble a full structure. For the design with a given metamaterial family, we can directly generate the corresponding microstructure by interpolating in the latent space as illustrated in Section 4.3. However, to design with the full database, we need to ensure that neighboring elements in the full structure are compatible on the shared boundary. As demonstrated in Section 4.2, a diverse candidate set can be obtained efficiently either by selection or optimization. The best-match pair is then selectively chosen from the candidate sets to achieve the designed properties while ensuring compatibility. Herein, we follow the method proposed in our previous paper [19], transforming the combinatorial problem into an inference process on a grid-like Markov random field (MRF). Specifically, unit cells in the full structure are mapped to the nodes of a grid-like undirected graph, with edges connecting adjacent unit cells. For each node, a diverse candidate set with $N_c$ microstructures can be obtained to meet the designed property by selecting from the database or via optimization with diverse initial guesses. A nodal weight $\theta_i(l_i)$ is defined for the $l_i^{\text{th}}$ microstructure candidate in the $i^{\text{th}}$ node to measure the deviation from the target property:

$$\theta_i(l_i) = \|\boldsymbol{C}_i - \boldsymbol{C}_{i,target}\|_\infty, \quad l_i = 1,2,\ldots,N_c, \quad (16)$$

where $\boldsymbol{C}_i$ and $\boldsymbol{C}_{i,target}$ are the real and target elastic stiffness of the $i^{th}$ node, respectively. For different combinations of microstructures $(l_i, l_j)$ of the neighboring unit cell pair $(i,j)$, an edge weight is assigned to the corresponding graph edge to measure incompatibility:

$$\theta_{ij}(l_i, l_j) = \theta_{ij}^G(l_i, l_j) + \theta_{ij}^M(l_i, l_j). \tag{17}$$

The first term is to measure the geometrical difference on the shared boundary, which is defined to be the ratio of incompatible binary elements to all solid elements on the shared boundary. The second term is to measure the mechanical incompatibility, which is defined to be the relative sum of stress difference on the shared boundary under the unit strain field. As illustrated in [28], by minimizing the second term, the overall performance predicted by the effective properties can remain relatively precise for aperiodic design.

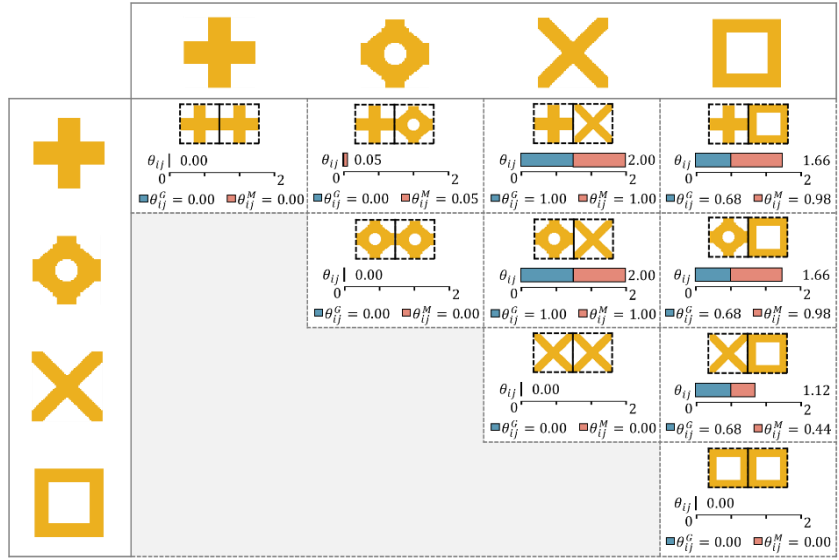

Fig. 14 Different pairs of horizontally connected microstructures and corresponding two measures of incompatibility (lower values preferred)

To provide a more intuitive illustration, we select several representative microstructures and report the geometric and mechanical incompatibility values below in the horizontal direction for all pairs, as shown in Fig. 14. Pairs located on the diagonal are composed of identical microstructures, which have the ideal compatibility with geometrical and mechanical measures $\theta_{ij}^G$ and $\theta_{ij}^M$ equal to 0. However, to achieve spatially varying properties, it's more common to have a neighboring pair with different microstructures, as shown in the off-diagonal regions. From the pairs in the third column, it can be noted that fully disconnected pairs will have the worst geometric and mechanical incompatibility measures. Meanwhile, as shown in the last column, for pairs with the same geometrical incompatibility measures (2nd and 3rd rows), their mechanical incompatibilities $\theta_{ij}^M$ can be quite different. Neighboring microstructures with a more consistent force transition path will have a lower (better) mechanical incompatibility measure. Therefore, the combination of geometrical and mechanical measures provides an effective quantification of the incompatibility between neighboring microstructures.

With these measures for the incompatibility, the original optimal full assembling problem is transformed into the label selection problem for the constructed weighted graph to minimize the sum of all the nodal and edge weights. To solve this graph optimization problem is equivalent to solve the inference problem on a grid-like Markov random field (MRF), whose optimal solution can then be found approximately but efficiently with the dual decomposition (DD-MRF) method [53].

**5.2 Multiscale system design examples**

In this section, the proposed two-stage design framework for the multiscale system is applied to both functionally graded and aperiodic structures.

The first example is to design a graded structure with the metamaterial families generated in Section 4. The setting of the design problem is given in Fig. 15(a). Specifically, the outer surfaces of the final design should form the target displacement curve in red, after squeezing the structure on the right end with a prescribed displacement. The design region for the beam is discretized into a 15×40 design mesh with each element to be filled by a microstructure from the given metamaterial families. In the macro-property optimization stage, we set the gradation of properties as an equality constraint in the optimization problem. The iterative optimization of the macroscale -property distribution is performed until the maximal change of the variables is less than 0.001, or until the number of iterations reaches 500. The optimized distribution for the $C_{11}$ values is shown in Fig. 15(b) while other components of the stiffness matrix can be obtained by Eq. (8). In the microstructure assembly stage, corresponding microstructures are generated by the decoder to assemble the full structure as shown in Fig. 15(c). After loading, the displacement profile precisely matches the design target with its relative root mean square error (RRMSE) to be 0.1900, as shown in Fig. 15(d).

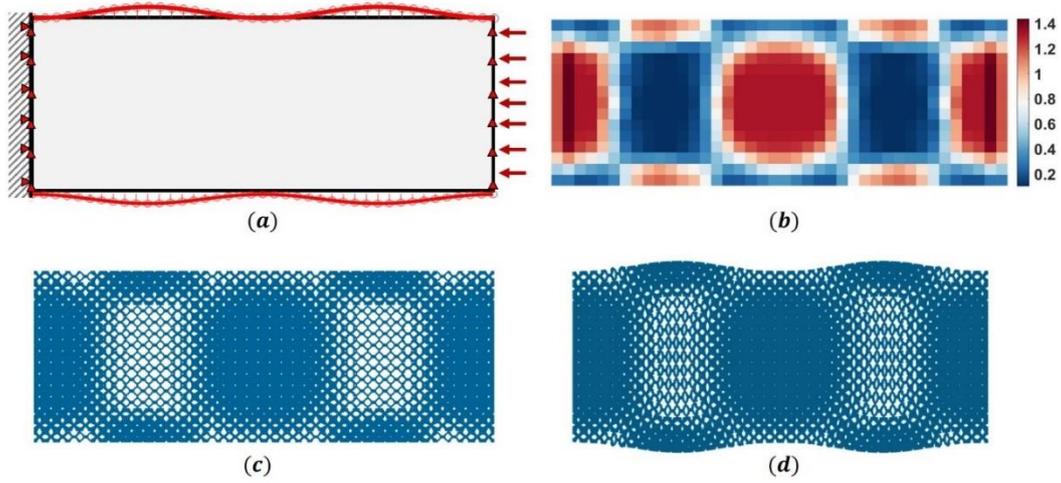

Fig. 15 The target displacements and design results for the first design case

When we need to achieve more intricate structural behavior, a larger property space is needed to enable a more versatile spatial property distribution. Therefore, in the following two examples, we turn to the design with the full metamaterial database. As shown in Fig. 16(a), the first example is to design a beam filled by spatially varying metamaterials, achieving a bridge-like displacement profile under horizontal compression with prescribed displacement. The design region for the beam is discretized into a $4 \times 30$ design mesh with each element filled by a $50 \times 50$ microstructure. Therefore, the total number of finite elements is 300,000 with 206 shared boundaries for adjacent microstructures. In the macro-property optimization stage, we use the signed L2 field as the inequality constraint of the properties. The macro property distribution is optimized with the same termination criteria for the graded structure design. With the optimized properties distribution, a diverse candidate set containing 10 microstructures is obtained for each element by using the microstructure selection and design methods proposed in Section 4. In the microstructure-assembly stage, the DD-MRF method is performed on the constructed weighted graph until convergence, i.e., the gap between primal and dual objective functions is closed, or the number of iterations exceeds 5000.

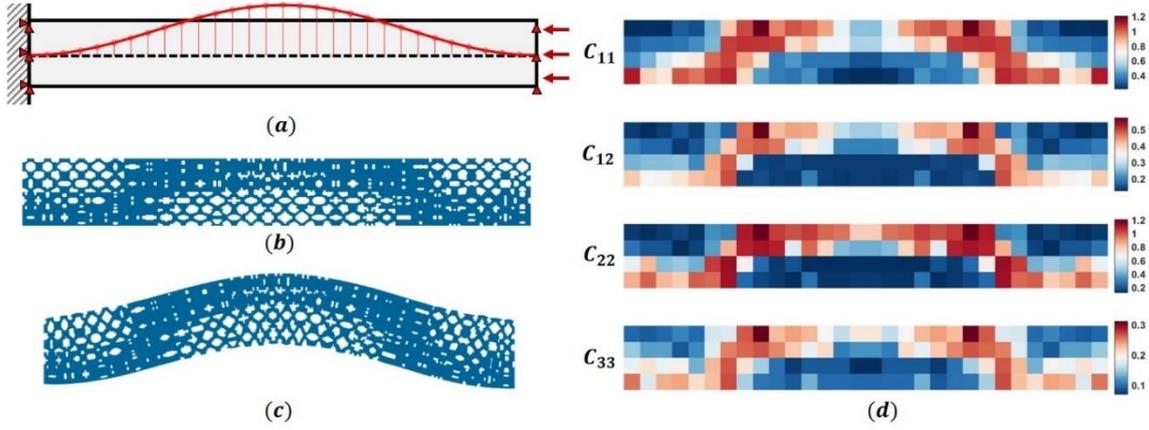

Fig. 16 The target displacements and design results for the second design case

The two-stage optimization is completed in 20 minutes with two CPUs (Intel i7-9750H 2.6GHz). The optimized distribution for different components of the stiffness matrix is shown in Fig. 16(d). We observe that different components of the stiffness matrix have similar spatial distributions, with stiffer materials distributed toward the upper part. It can be expected that the lower half of the beam will have a larger compression than the upper half, resulting in upward bending. By taking advantage of the dense dataset and the highly structured latent space, compatible microstructures can be obtained from the diverse candidate set for each unit cell. This is indicated by the full structure design shown in Fig. 16(b), where each pair of adjacent unit cells is well connected on the shared boundary. Due to the optimized properties distribution and the compatible boundaries, the assembled structure achieves a bridge-like shape that is well-matched with the design target (RRMSE = 0.0600), as shown in Fig. 16(c).

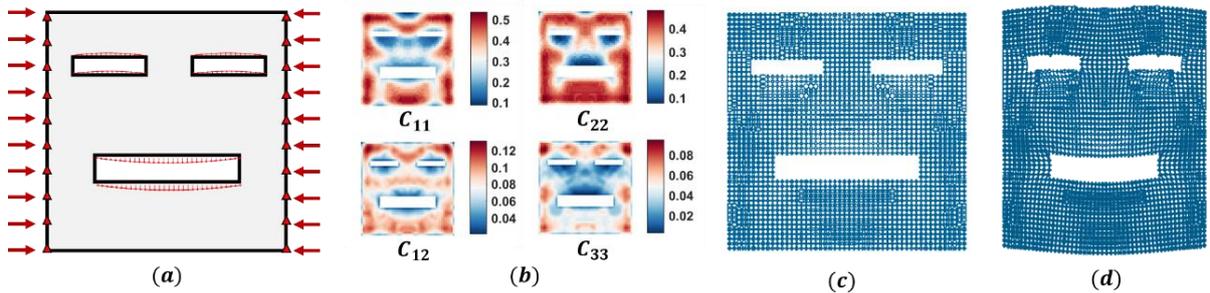

Fig. 17 The target displacements and design results for the third design case

To further demonstrate the ability of the proposed method, a more complicated design problem is given in Fig. 17(a). Here we aim to activate a smiley face by squeezing the left and right ends of the structure. Compared with the previous examples, this design includes finer discretization ($50 \times 50$ design mesh with 6.25 million finite elements) with more shared boundaries and a relatively distinct but coupled displacement profiles in different regions. The whole optimization process is completed in 11 hours and 12 minutes on 6 processors (Intel i7-9750H 2.6GHz). This computation time is satisfying for a full structure with 6.25 million elements and nearly 4900 shared boundaries. Since the DD-MRF method is highly amenable to parallel computation with the dual decomposition, the computation cost can be further reduced by using more processors.

The optimized distributions of macro-properties are shown in Fig. 17(b), where different components of the stiffness matrix have relatively different distribution patterns. This requires numerous microstructures with different anisotropic characteristics, which benefits from our large metamaterial database. Despite the complexity of the design problem, Fig. 17(c) and (d) demonstrates that our method still obtains a full structure with compatible boundaries. The heterogeneous property distribution is achieved by aperiodically assembling different metamaterial microstructure designs, causing a smiley face to form upon loading (RRMSE = 0.1623). For the conventional single scale TO method or other aperiodic multiscale design methods, the design process would be more computationally demanding due to a large number of design variables and numerous constraints on the shared boundaries. We also observed that a single-scale SIMP method will tend to render unrealistic designs with numerous intermediate density values for this design problem. A possible explanation for this is that, under the prescribed displacement boundary condition, the objective function only depends on the relative distribution rather than the absolute property values, which is different from the design with force boundary condition. This will paralyze the SIMP scheme and result in the infeasible design. In contrast, our design process directly uses the properties as design variables in the first stage, which is independent of the geometric representation of the solid structure. This is another advantage provided by the proposed data-driven method.

In this study, we only focus on the design for target displacement profile in this study instead of simpler compliance minimization, which can be view as a special case of target displacement design with extra volume constraint. This is because structures with porous materials are reported [54, 55] to be sub-optimal and offer little benefit for compliance minimization design that only pursues the overall stiffness of the structure. As a result, the potential of a large database with versatile properties cannot be fully exploited. In contrast, multiscale structures have shown to have the edge over single scale structure for the design involving multi-physics or spatially varying structural requirements, such as displacement design [23, 28], thermal-elastic problem [56, 57], dynamic response optimization [58, 59] and energy absorption design [60]. The results of displacement designs shown in this study represent an encouraging initial step toward the further application of metamaterials in these promising areas.

## 6. Conclusions

A scalable integrated framework is proposed for the data-driven design of metamaterials in multiple scales by taking advantage of the continuous and highly structured latent space of the VAE model. We highlight that a well-trained VAE model that is integrated with a structure-property regressor can distill salient geometrical features from metamaterial microstructures to form a continuous and meaningful latent space. Complex shape transformation is encoded in different moving directions in the latent space, rendering a natural interpolation method and a distance metric to measure shape similarity. Semantic structures of the latent space are identified to enable higher-level control of the mechanical properties. These mechanistic insights provide an efficient representation and management structure for the microstructure design, metamaterial family generation, and the assembling of multiscale metamaterial systems.

Through microstructure design, we demonstrated that complex mappings on the topology and mechanical properties of the microstructures can be easily achieved by simple vector arithmetic in the latent space. The distance metric induced by the latent space can be used to select or generate a diverse candidate set for a given property by clustering in the latent space. For the design of metamaterial families, we proposed to generate a diverse set of metamaterial families with target graded properties by searching on a directed graph constructed in the latent space. The design methods at the microstructure and family levels are integrated into a two-stage framework for multiscale system design to achieve the prescribed distortion of the full structure, ensuring the compatibility between adjacent unit cells. Greater computational efficiency is obtained by replacing the nested optimization with a precomputed database. The same database and design algorithm can be directly applied to efficiently design multiscale systems with different loading conditions or objective functions.

Since the meaningful latent space is a result of its continuity and low dimensionality, we believe that mechanistic insights obtained in this study have the potential to be extended to general microstructural materials. To apply the proposed framework to 3D cases, future work needs to be carried out to enable the VAE construction of 3D microstructures with advanced machine learning techniques, e.g., using 3D voxels or point cloud representations. Also, since the homogenized property calculation becomes rather costly for 3D metamaterials, a more sophisticated database construction method is necessary. We have ongoing research in this direction [61], including avoiding bias towards particular properties or shapes that contribute little to the quality of property and shape exploration. Currently, our work relies on the homogenization theory and only focuses on the design in linear elasticity. For future works, we will explore methods to address the aperiodic boundary condition and the scale-related issue. Applying the proposed data-driven design framework to thermal-elastic design, dynamic response optimization, and some other multi-physics problems is underway. Also, imposing manufacturing constraints on the current framework is another promising and important future direction to explore. Complex manufacturability constraints can be incorporated into our learning and optimization approaches by either implicitly modifying the loss function for the training process or using an explicit penalty in the later design stage.

**Acknowledgments**

We are grateful for support from the NSF CSSI program (Grant No. OAC 1835782). Liwei Wang acknowledges support from the Zhiyuan Honors Program for Graduate Students of Shanghai Jiao Tong University for his predoctoral visiting study at Northwestern University. Yu-Chin Chan thanks the NSF Graduate Research Fellowship (Grant No. DGE-1842165).

# Appendix A

This part presents the detailed sensitivity analysis of the full structure optimization problem proposed in Section 5.

We rewrite the objective function to add the equilibrium constraint:

$$\mathcal{F}(\boldsymbol{C}_e) = \|\boldsymbol{\gamma}\odot\boldsymbol{u} - \boldsymbol{u}_t\|_2^2 - \boldsymbol{\lambda}^T(\boldsymbol{K}(\boldsymbol{C}_e)\boldsymbol{u} - \boldsymbol{f}), \tag{A1}$$

where $\boldsymbol{\lambda}$ is an arbitrary real vector serving as the multiplier. Taking the derivative with respect to $C_{e,ij}$ on both sides of the equation, we can obtain

$$\begin{aligned}\frac{\partial \mathcal{F}}{\partial C_{e,ij}} &= 2(\boldsymbol{\gamma}\odot\boldsymbol{u} - \boldsymbol{u}_t)^T\boldsymbol{\gamma}\odot\frac{\partial \boldsymbol{u}}{\partial C_{e,ij}} - \boldsymbol{\lambda}^T\left(\frac{\partial \boldsymbol{K}}{\partial C_{e,ij}}\boldsymbol{u} + \boldsymbol{K}\frac{\partial \boldsymbol{u}}{\partial C_{e,ij}}\right)\\ &= 2[\boldsymbol{\gamma}\odot(\boldsymbol{u}-\boldsymbol{u}_t)]^T\frac{\partial \boldsymbol{u}}{\partial C_{e,ij}} - \boldsymbol{\lambda}^T\boldsymbol{K}\frac{\partial \boldsymbol{u}}{\partial C_{e,ij}} - \boldsymbol{\lambda}^T\frac{\partial \boldsymbol{K}}{\partial C_{e,ij}}\boldsymbol{u}\\ &= \{2[\boldsymbol{\gamma}\odot(\boldsymbol{u}-\boldsymbol{u}_t)]^T - \boldsymbol{\lambda}^T\boldsymbol{K}\}\frac{\partial \boldsymbol{u}}{\partial C_{e,ij}} - \boldsymbol{\lambda}^T\frac{\partial \boldsymbol{K}}{\partial C_{e,ij}}\boldsymbol{u}.\end{aligned} \tag{A2}$$

This can be transformed to be

$$\frac{\partial \mathcal{F}}{\partial C_{e,ij}} = -\boldsymbol{\lambda}^T\frac{\partial \boldsymbol{K}}{\partial C_{e,ij}}\boldsymbol{u}, \tag{A3}$$

when $\boldsymbol{\lambda}$ satisfies the following adjoint equation:

$$2[\boldsymbol{\gamma}\odot(\boldsymbol{u}-\boldsymbol{u}_t)]^T - \boldsymbol{\lambda}^T\boldsymbol{K} = \boldsymbol{0}. \tag{A4}$$

The solution for this adjoint equation can be obtained as

$$\boldsymbol{\lambda}^T = 2[\boldsymbol{\gamma}\odot(\boldsymbol{u}-\boldsymbol{u}_t)]^T\boldsymbol{K}^{-1}. \tag{A5}$$

Therefore, the derivative of the objective function with respect to the component of the stiffness matrix can be found as

$$\frac{\partial \mathcal{F}}{\partial C_{e,ij}} = -2[\boldsymbol{\gamma}\odot(\boldsymbol{u}-\boldsymbol{u}_t)]^T\boldsymbol{K}^{-1}\frac{\partial \boldsymbol{K}}{\partial C_{e,ij}}\boldsymbol{u}. \tag{A6}$$

From Eq.(9), we can further obtain:

$$\frac{\partial \mathcal{F}}{\partial C_{e,ij}} = -2[\boldsymbol{\gamma}\odot(\boldsymbol{u}-\boldsymbol{u}_t)]^T\boldsymbol{K}^{-1}\frac{\partial \widetilde{\boldsymbol{K}}_e}{\partial C_{e,ij}}\boldsymbol{u}. \tag{A7}$$

At the element level, $\frac{\partial \boldsymbol{K}_e}{\partial C_{e,ij}}$ can be expressed as

$$\frac{\partial \boldsymbol{K}_e}{\partial C_{e,ij}} = \boldsymbol{B}^T\frac{\partial \boldsymbol{C}_e}{\partial C_{e,ij}}\boldsymbol{B}. \tag{A8}$$

This expression can be transformed into the global form and then put into (A7) to obtain the sensitivity value.